\documentclass{PoS}
 \usepackage{multicol}
\usepackage{floatrow}
\usepackage{wrapfig}
\newfloatcommand{capbtabbox}{table}[][\FBwidth]

\usepackage{blindtext}

\title{Atmospheric pressure dependance of HAWC scaler system}

\ShortTitle{Pressure dependance on HAWC scaler system}

\author{{\speaker{K.P. Arunbabu$^1$}, Alejandro Lara$^{1,2}$, James Ryan$^3$ for the HAWC collaboration}\\
        $^1$ Instituto de Geof\'{i}sica, Universidad Nacional Aut\'{o}noma de M\'{e}xico, Mexico City, Mexico\\
        $^2$The Catholic University of America, USA\\
        $^3$ University of New Hampshire, USA.\\
        E-mail: \email{arun@igeofisica.unam.mx}\\  
	  \footnote{for collaboration list see PoS(ICRC2019)1177} For a complete author list and acknowledgement see  \\
	{\href{https://www.hawc-observatory.org/collaboration/icrc2019.php}{$https://www.hawc-observatory.org/collaboration/icrc2019.php$}} 
        }

\abstract{The variation in atmospheric pressure is due to changes in mass of the air column above, which in turn resembles the density variation of atmosphere and will affect the decay of secondary particles of cosmic rays. The ground based cosmic ray detectors observe pressure dependent variation in their flux. The High Altitude Water Cherenkov (HAWC) gamma ray observatory is a great detector of secondary particles because of its high altitude, high uptime, and large area (including total photo-cathode area), which makes the HAWC scaler system an ideal instrument for solar modulation studies. Although, in order to perform these studies it is necessary to isolate and remove the atmospheric modulations. The observed rate in each PMT has signatures of both the solar and atmospheric modulations, which makes it difficult to measure the pressure coefficient ($\beta_P$). The pressure at the HAWC site shows a periodic behavior ($\sim$ 12 hours), which also reflects in the scalar rates. This periodic property was used to isolate the pressure modulation and $\beta_P$ were estimated with accuracy. Since the pressure dependence is a physical phenomenon, the estimated coefficients for PMTs should be identical, any deviation from this can be due to malfunction of the PMT. This make this method a useful tool to identify the malfunctioning PMTs and help us to isolate them from the analysis. In this analysis we are presenting the method of estimation of the pressure coefficients and its usage to correct the HAWC scalar data to make it suitable for the solar modulations studies. }

\FullConference{36th International Cosmic Ray Conference -ICRC2019-\\
		July 24th - August 1st, 2019\\
		Madison, WI, U.S.A.}

\begin{document}

\section{Introduction}

Modulations in the intensity of  galactic cosmic ray (GCR),  observed using the ground-based instruments  have been studied for decades \cite{Forbush54,bur85,Prasad09,Arunbabu13,Arunbabu15}. These variations can be caused by  solar activity or due to the effect of Earth's atmosphere. The solar modulations of GCRs observed at Earth have served as  a useful  tool to study the space-weather effects caused by  transient phenomena such as  solar flares, coronal mass ejections (CMEs), and coronal holes.  When the GCRs interact with the atmosphere, they produce secondary particles that ground-based detectors can measure. These measurements can be used to study the solar modulations of GCRs, which are mainly observed in the secondaries produced by the primary cosmic rays of energies $<$100\,GeV.  Also the secondaries produced in the atmosphere will get modulated by the atmosphere, which depends on parameters such as pressure and temperature \cite{pkm16,aru17}. The rarer the atmosphere the lesser will be the interaction of particles, hence more decay, whereas for a denser atmosphere the decay rate will be less. 
It is essential to identify these atmospheric modulations due to pressure and temperature and make the corresponding corrections  to use secondary particle data for solar modulation studies. In this work, we explain the methods used to  identify and correct the pressure modulations in the HAWC TDC-scaler system.  

\begin{wrapfigure}{r}{0.45\linewidth}
\centering
\includegraphics[width = 0.9\linewidth]{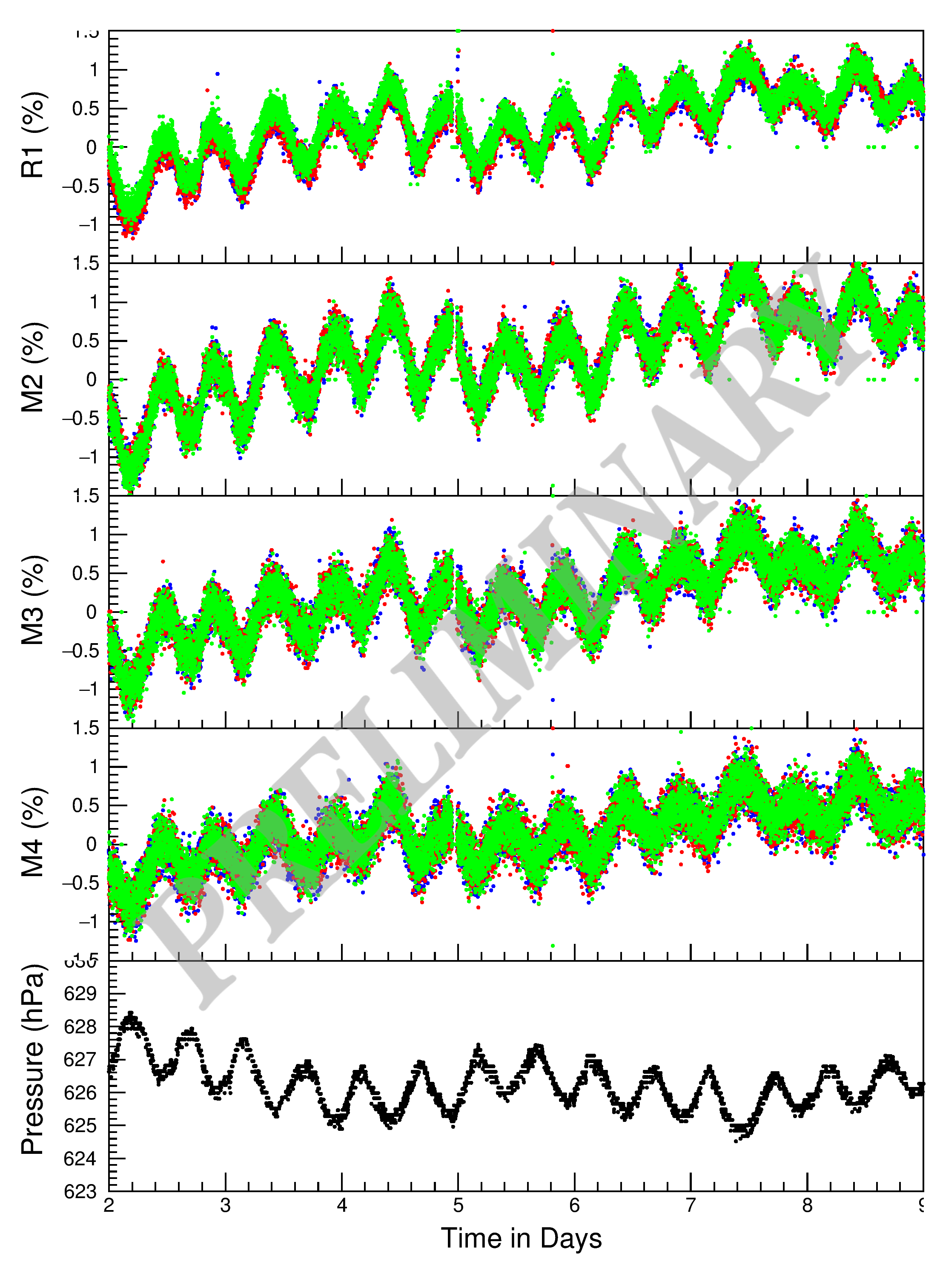}
 \caption{{\scriptsize  \it \label{fig1}The top four panels show the observed HAWC TDC-scaler data $R_{1}$ and the multiplicity rates $R_{M2}$, $R_{M3}$, $R_{M4}$  respectively from top to bottom.  Rates of a few example PMTs and tanks are shown in  different colors. The bottom-most panel shows the ambient pressure at HAWC site.} }
\end{wrapfigure}

The GCRs reaching the top of the atmosphere interact with   atmospheric nuclei and produce an increasing flux of secondary particles as they propagate downwards. These secondary particles mainly consist of neutrons and mesons (pions and kaons). Due to the relatively long lifetime ( $\sim$15 min) of neutrons,  a good fraction of them survives down to the ground-level, whereas the mesons decay because of their short lifetime and produces muons. The majority of these muons survive down to the ground-level affected by an energy-loss mechanism that is dominated by ionization. A large fraction of these  muons is produced higher up in the atmosphere at $\ge$ 6 km \cite{aru17}, and constitute the dominant fraction of secondary cosmic rays at sea-level.  

\section{HAWC Data sets}

The High Altitude Water Cherenkov (HAWC) observatory is located on a relatively flat piece of land near the saddle region between the Sierra Negra and Pico de Orizaba, with latitude $18^{\circ}59' 41"$ N,  longitude $97^{\circ} 18'30.6"$ W and altitude at 4100 m above sea level. HAWC consists of 295 water Cherenkov detectors (WCD) spread over an area of 20,000 $m^2$, each of it is  7.3 m in diameter and 4.5 m in depth. Each of these WCDs is filled with filtered water and instrumented with  4 photomultiplier tubes (PMTs). A 10-inch PMT at the center of the WCD is at positions `C', and three 8-inch PMTs are arranged around the central one making an equilateral triangle of side 3.2 m at positions `A, B and D'.  The TDC-scaler system of HAWC records the output of each one of the 1180 PMTs (R1) as well as  the multiplicity rates $M_2$, $M_3$ and $M_4$ from each WCD. In this analysis we will be using data from this TDC-scaler system. This analysis is carried out using  `1-minute'  averaged data from the HAWC TDC-scaler system, for the months of September, October and November of the year 2016.  We  used data from all 1180 PMTs and multiplicities $M_2$, $M_3$ and $M_4$, which will be called  $R_{1}$, $R_{M2}$, $R_{M3}$ and $R_{M4}$ respectively. 
The vertical cut-off rigidity of HAWC is 7.9 GeV \cite{lar13} and the median energy rigidity is 41.97, 41.46, 42.28, and 45.04 GeV respectively for $R_1$, $M_2$, $M_3$ and $M_4$. The atmospheric pressure at the HAWC site is measured  every minute using a barometer. 
The variation of $R_{1}$, $R_{M2}$, $R_{M3}$, $R_{M4}$  and the atmospheric pressure over a period of 7 days from 2 to 8 October 2016 are shown in figure \ref{fig1}. An anticorrelation between atmospheric pressure and TDC-scaler rates are clearly visible in this figure and a dominant 12 hour periodicity in both the data sets are also observed.  As HAWC used PMTs of different sizes, e.g., 8-inch and 10-inch, the mean rates of PMTs ($<Rm_{1}>$) are different and the rates also  depend upon each PMTs gain, operation voltage, and its quantum efficiency. Since the mean rates are spread over a large range ($\sim 400  - 800$ counts/sec), we consider the percentage variation ($\frac{R_{1}-<Rm_{1}>}{<Rm_{1}>} \times 100$) of rates for each PMTs and multiplicities in our analysis. Since the  physical phenomena of  solar and atmospheric modulations cover a large area and affect the PMTs uniformly, the percentage variations observed in each PMT will have the same percentage variations irrespective of their gain.

\section{Estimation of pressure coefficient {$\beta_P$}}

The variation in atmospheric pressure is due to the changes in the mass of the air column above the detector which in turn results in a corresponding variation in the flux of secondary particles. This effect is observed from the anti-correlation of the TDC-scaler rate and its multiplicities with the pressure as shown in figure \ref{fig1}. The secondary particle rate observed using the ground-based detectors is affected by the modulations due to solar origin along with the atmospheric origin, which makes the estimation of pressure or temperature dependence of the secondary particle rate more difficult \cite{pkm16,aru17}. Due to its near-equatorial location ($18^{\circ}N$) the pressure at the HAWC site shows a periodic variation with a period of $\sim$ 12 hrs \cite{lin79,car09},  which is a tidal effect.  This is due to solar heating function, what we see at HAWC is the effect of a westward propagating gravity wave that is Sun synchronous.  Basically the Sun heats the atmosphere and its scale height increases, then gravity pulls the gas back down.  The solar heating function is like a square wave with many harmonics, so the second harmonic can be stronger than the fundamental. The observed TDC-scaler rate also show a synchronous periodic responses in anti-correlations with the pressure variations.  We used this periodic behavior to estimate the pressure dependence in the TDC-scaler rate using the `Fast Fourier Transform´ (FFT) on both the data sets. 

\begin{figure}
\begin{floatrow}
\ffigbox{%
 \includegraphics[width = 0.8\columnwidth]{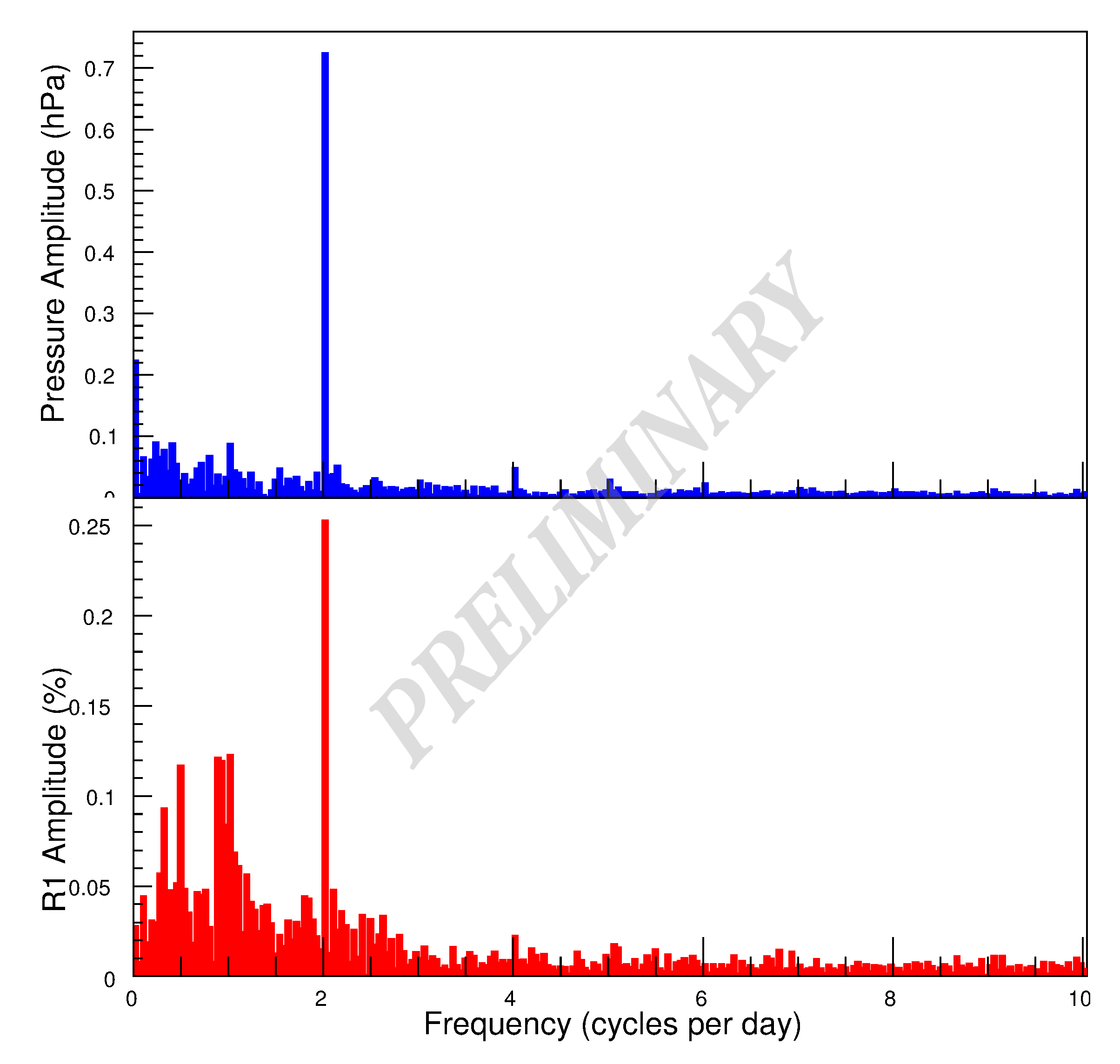}
}{%
 \caption{{\scriptsize  \it \label{pwrs} FFT spectrum of \textcolor{blue}{pressure} at the top panel and \textcolor{red}{TDC scaler rate $R_1$} at bottom panle, for the month of October 2016.} }
}
\ffigbox{%
 \includegraphics[width = 0.8\columnwidth]{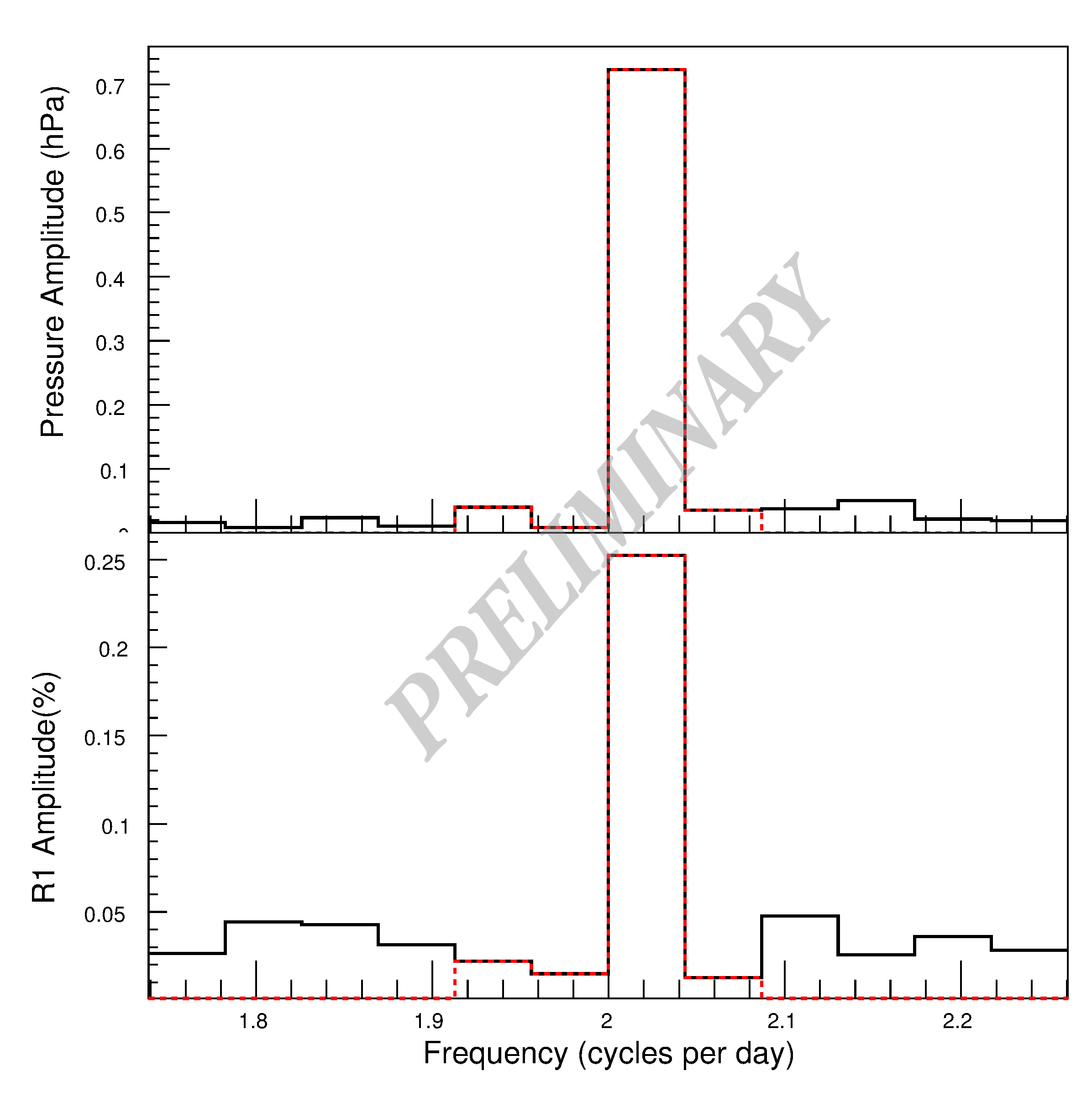}
}{%
\caption{{\scriptsize  \it \label{fps} power spectrum of pressure (top panel) and TDC scaler rate $R_1$ (botom panel). Filtered spectra are shown by red dashed line and the original one is shown by the black solid line.  }}
}
\end{floatrow}
\end{figure}

The FFT power spectra for the scaler rates  $R_{1}$ and multiplicity rates $R_{M2}$, $R_{M3}$, $R_{M4}$ and pressure were calculated. Examples of the power spectra of $R_{1}$  and pressure are shown in figure \ref{pwrs}. 
 It is clear from the figure that the pressure spectra has a dominant peak around 2 cpd and the rates $R_{1}$ also show a dominant peak corresponding to this frequency. The power spectra of TDC-scaler rates $R_{1}$   show another dominant peak which is possibly due to the solar modulation present on the TDC-scaler rate such as the 1 cpd frequency that corresponds to the solar diurnal anisotropy. The dominant peak of the power spectra of pressure and TDC-scaler rates at 2 cpd implies a significant contribution from pressure variation on the TDC-scaler rate observed. This feature in the data sets was exploited to segregate the non-barometric effects from the TDC-scaler rates $R_{1}$, $R_{M2}$, $R_{M3}$,\& $R_{M4}$ and used to estimate an accurate pressure coefficient ($\beta_P$) for these  rates. To extract only the 2 cpd barometric effects from the data sets we used a narrow-band filter W(f). Similar filters were used in the past to extract the atmospheric effects from muon variations observed in another muon detector \cite{pkm16,aru17}. The filter was designed to select the frequencies centered at 2 cpd and is described as below,

{\scriptsize \begin{eqnarray}
W(f) = \left\{
\begin{array}{l l l} 1, \;\;\;\;\;\;\;\;\;\;\;\;\;\;\ \;\;\;\;\;\; \textit{if}\;\; |f-f_{c}| \leq \Delta f \\
\sin \frac {\pi} {2}  \frac{|f - f_{c}|}{\Delta f}, \;\;\;\;\; \textit{if}\;\; \Delta f < |f - f_{c}| \leq 2 \Delta f \\
0, \;\;\;\;\;\;\;\;\;\;\;\;\;\;\;\;\;\;\;\;\;\;  \textit{if}\;\; |f - f_{c}| > 2 \Delta f
\end{array} \right.
\end{eqnarray}}
Here the central frequency is represented by $f_c$, in our  analysis  the filter was constructed with $f_c=2$ $ cpd$ and $\Delta f$ = 0.01. This filter has a 100\% acceptance within the range from 1.99 to 2.01 cpd and the acceptance is gradually decreases to zero following a sinusoidal behavior in the range from 1.99 to 1.98 cpd, and 2.01 to 2.02, respectively. Outside of this range of frequencies, the acceptance become zero. 

\begin{figure}[h]
\begin{floatrow}
\ffigbox{%
 \includegraphics[width = 0.95\columnwidth]{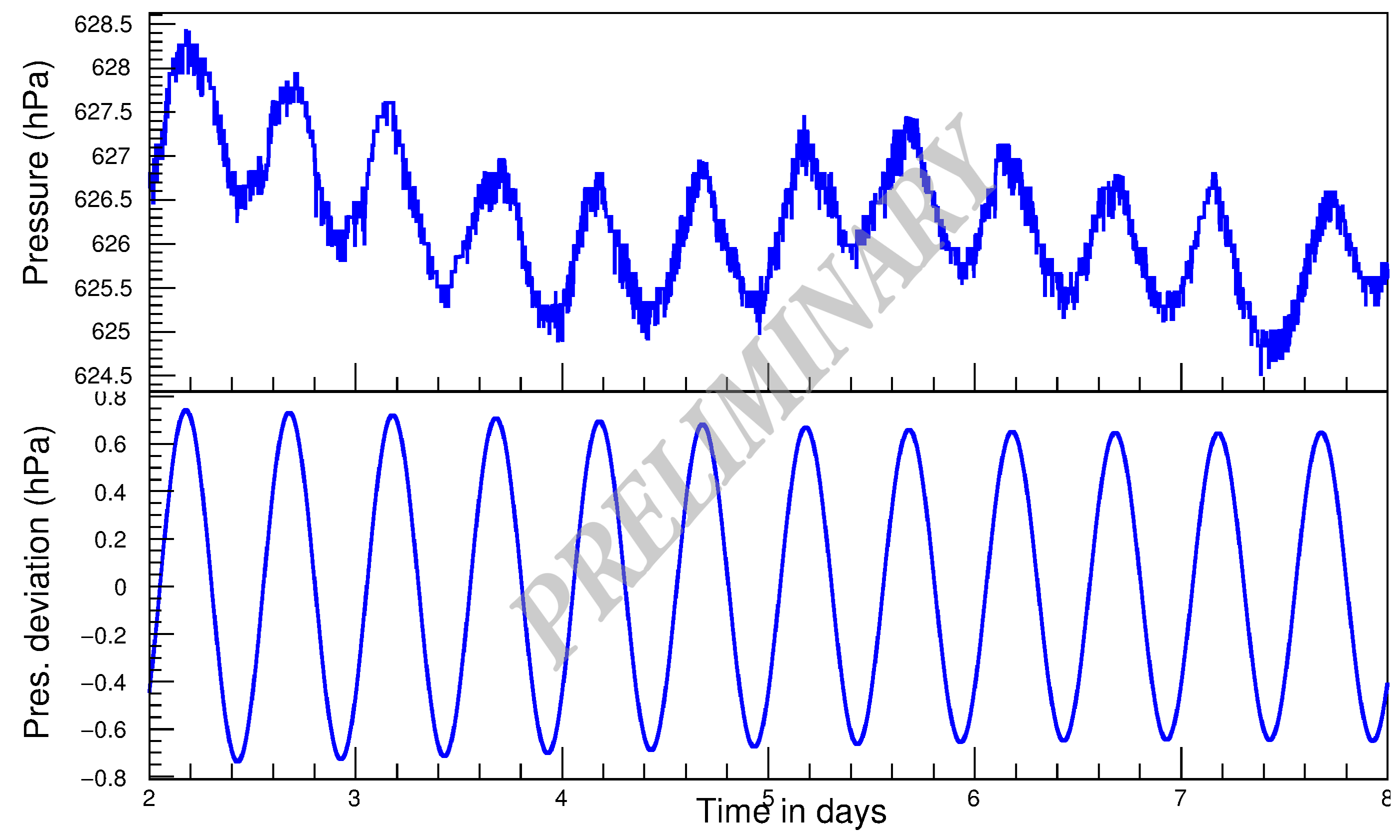}
}{%
 \caption{{\scriptsize  \it \label{filtp} Top panel shows the pressure variation at HAWC site from 2 October 00 hr from 8th October 00 hr, 2016. the bottom panel shows the IFT of the 12 hour periodic nature of pressure during the same time.} }
}
\ffigbox{%
 \includegraphics[width = 0.95\columnwidth]{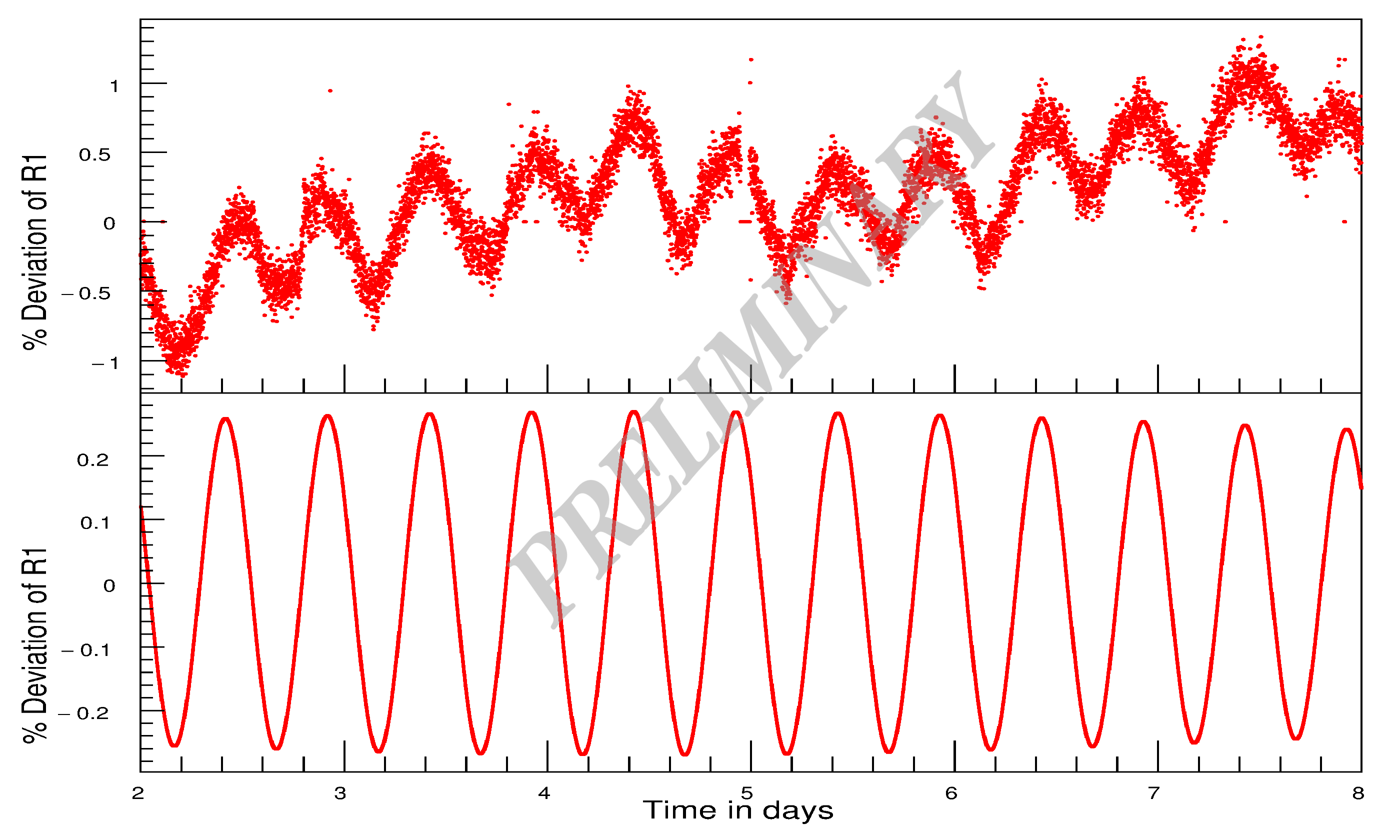}
}{%
\caption{{\scriptsize  \it \label{filtr} Top panel shows the TDC scaler rate $R_1$ variation observed  from 2 October 00 hr from 8th October 00 hr, 2016. and the bottom panel shows the IFT of the 12 hour periodic nature of  $R_1$   }}
}
\end{floatrow}
\end{figure}

\begin{figure}[h]
\begin{floatrow}
\ffigbox{%
\includegraphics[width = 0.75\columnwidth]{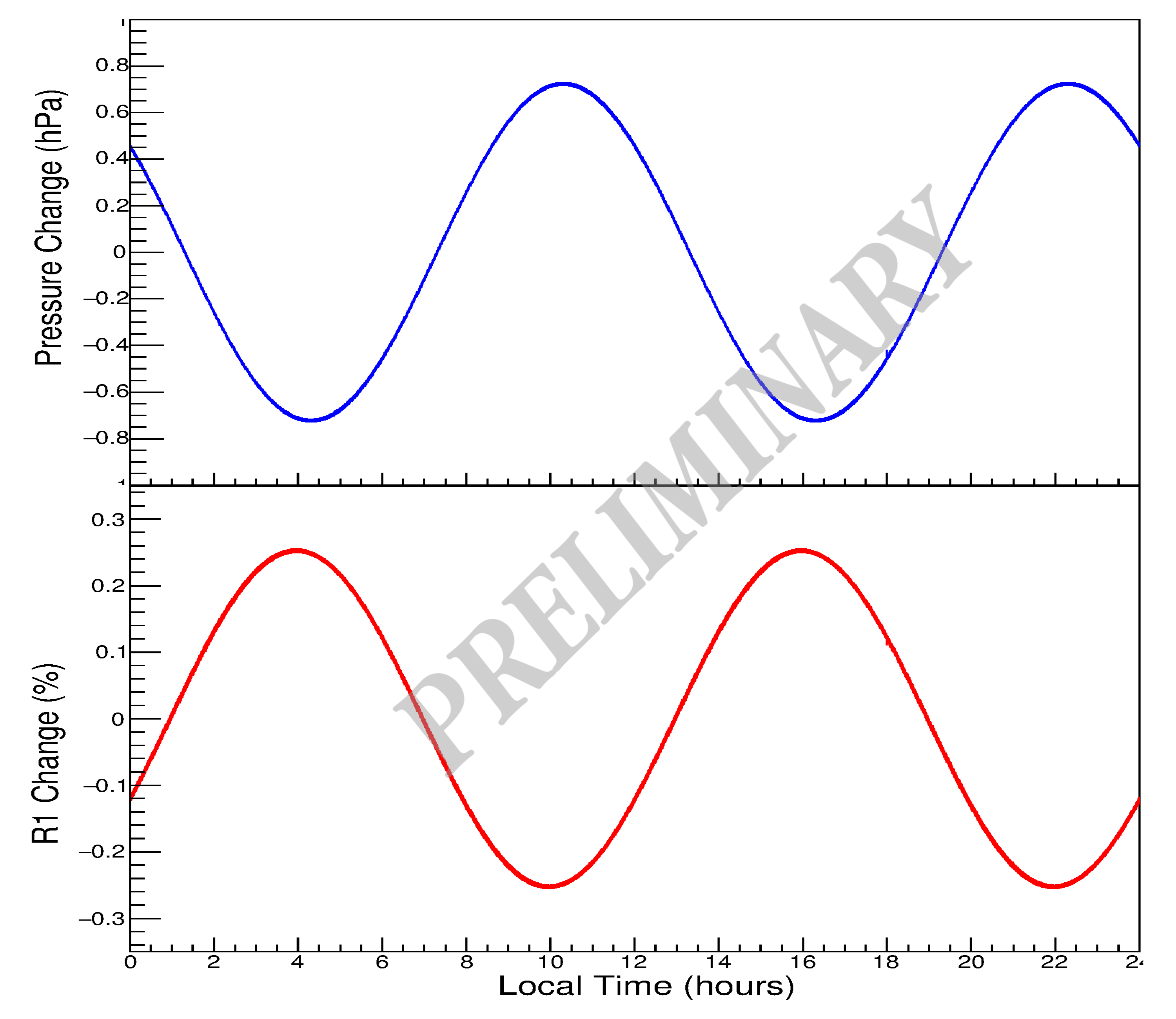}
 
}{%
\caption{{\scriptsize  \it \label{fpr} IFFT data of pressure and $R_1$ in local time domain, folded to a 24 hour format, top panel is of pressure and bottom one for TDC scaler rate $R_1$. } }
}
\ffigbox{%
 \includegraphics[width = 1.0\columnwidth]{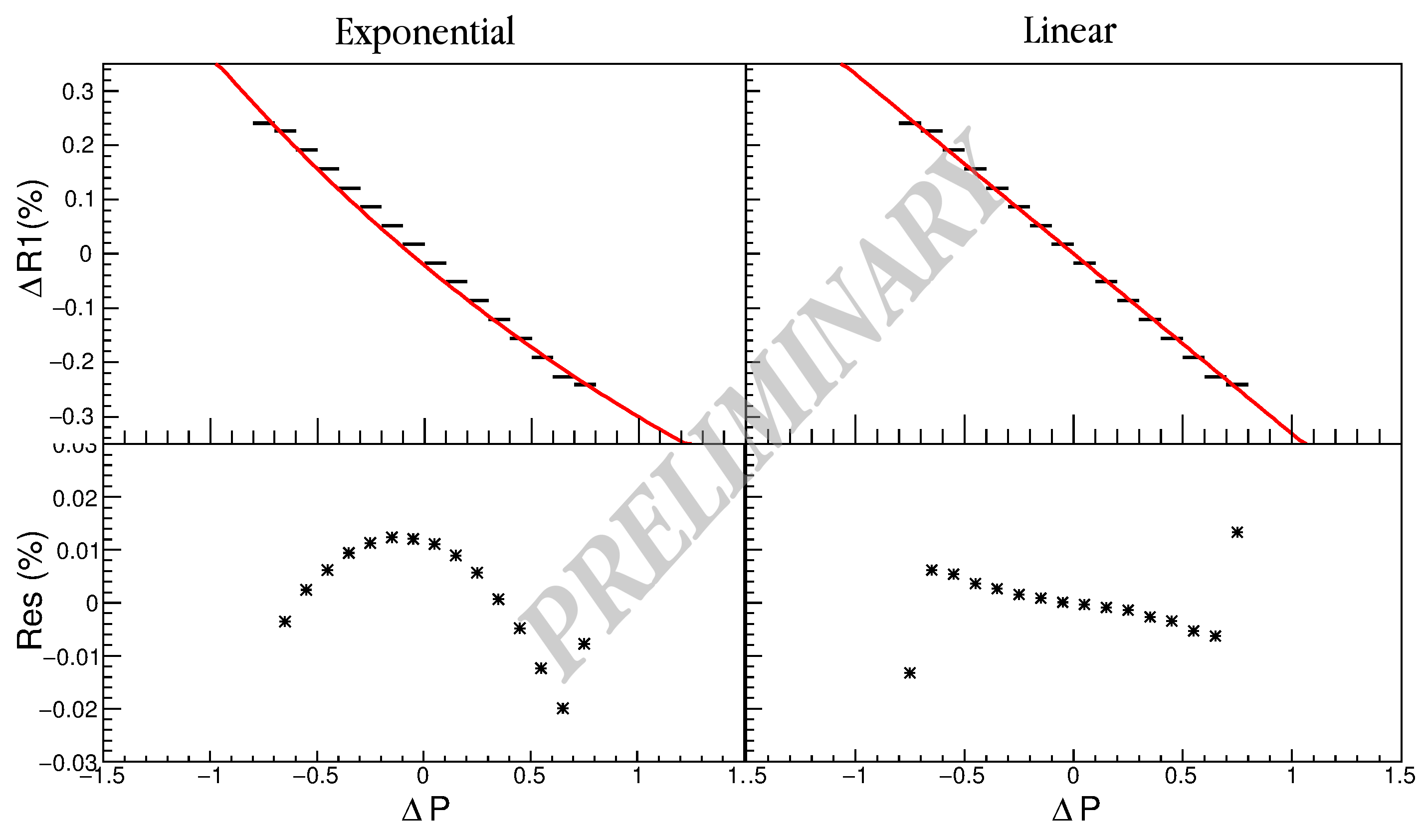}
}{%
\caption{{\scriptsize  \it \label{fit} Top panels shows the depndance of $R_1$ on pressure. First panel shows the exponential fit and the second panels shows the linear fit. Bottom panels shows the residual of fit to data.}   }
}
\end{floatrow}
\end{figure}

This narrow-band filter  was applied to the FFT frequency spectrum of the pressure and the TDC-scaler rates $R_{1}$, $R_{M2}$, $R_{M3}$, \& $R_{M4}$ . The resultant spectra contain only frequencies from 1.98 to 2.02 cpd, example is shown in figure  \ref{fps}. The filtered spectrum has a smooth sinusoidal transition on either side of $f_c$, and removes all the frequencies below 1.98 cpd and above 2.02 cpd. This filtered power spectra were converted back into the time domain by applying an inverse fast Fourier transform (IFFT).  The original pressure data and the IFT after filtering are shown in figure \ref{filtp}, and the same for $R_{1}$ are shown in figure \ref{filtr}. From the figures, we can see that the higher and lower order frequencies are entirely removed and the IFT data  has a periodicity of $\sim$ 12 hours, with the baseline of zero. The IFT data of these data sets were folded to fit into a 24-hour window and are shown in figure \ref{fpr}. The  `X' axis of this figure is given in the local time of Mexico, where we can see that the minimum of the pressure occurs at approximately 4 AM and 4 PM, whereas the maximum occurs at approximately 10 AM and 10 PM showing the 12 hour periodic nature \cite{car09}. A near perfect anti-correlation of pressure and  $R_{1}$ is also visible in this figure.

The dependence of $R_{1}$ on the atmospheric pressure can be found by plotting the IFFT of $R_{1}$ against the IFFT pressure data as shown in top panels in figure \ref{fit}. Each data point in these plots represents the mean $R_{1}$ for a pressure bin of width 0.1 hPa.  Empirically the dependence of TDC-scaler rates on the atmospheric pressure can be described by an exponential function as 

{\scriptsize
\begin{equation}
R (P) = R(P_m) ~~exp^{\beta_P \Delta P} \label{eq1}
\end{equation}}

where $R(P)$ is the rate observed at the ground based detectors at pressure P,  $R(P_m)$ is the rate at the mean pressure level, $\Delta P$ is  the deviation in pressure from its mean value $P_m$, and parameter $\beta_P$ is the pressure coefficient. It can be observed from the figures \ref{fps}, \ref{filtp} and \ref{fpr} that the amplitude of pressure is small and of the order of $\sim 0.7 hPa$.  Considering this value is less than 1, we took the first order linear approximation in Taylor series expansion of the exponential function in Eq.\ref{eq1} as, 

{\scriptsize
\begin{equation}
R (P) = R(P_m) ~~(1+{\beta_P \Delta P}) \label{eq2}
\end{equation}}

The HAWC site experiences stormy weather. Storms cause significant pressure variations, so in our analysis we carry out both the exponential and the linear approximation aproaches. The dependence of TDC-scaler rates $R_{1}$ on the atmospheric pressure are shown in figure \ref{fit}. The top left panel shows a fit of an exponential function ` $C_1 + exp^{\beta_P \Delta P}$'. In the top right panel we fit the linear approximation `$C_2+{\beta_P \Delta P} $'.  The bottom panles shows  the residual of fitting to the data. 

\begin{table}
{\footnotesize
 \caption{{\scriptsize  \it \label{tbeta} Mean value of $\beta_P$ for the months of September, October and November. }}
 \begin{tabular}{l|cc|cc} \hline \hline 
{Month } & {$\beta_P$  (\%/hPa)} &{$\beta_P$ (\%/hPa)}& {$\beta_P$  (\%/hPa)} &{$\beta_P$ (\%/hPa)} \\ 
&{(exponential)} & { (linear)} &{(exponential)} & { (linear)} \\ \hline  \hline
& \multicolumn{2}{c}{PMT rate R1} & \multicolumn{2}{c}{Multiplicity M2} \\   \hline 
{September } & {-0.3366$\pm$0.0004}   &  {-0.3433$\pm$0.0004} &    {-0.4161$\pm$0.0006}   &      {-0.4304$\pm$0.0007}      \\
{October  }&    {-0.3383$\pm$0.0004}   &{-0.3430$\pm$0.0004} &     {-0.4117$\pm$0.0004}   &   {-0.4200$\pm$0.0006 }   \\
{November }&  {-0.3358$\pm$0.0006}   & {-0.3416$\pm$0.0006}&   {-0.4015$\pm$0.0006}   &{-0.4112$\pm$0.0006}    \\  \hline \hline 
& \multicolumn{2}{c}{Multiplicity M3} & \multicolumn{2}{c}{Multiplicity M4}\\ \hline
{September }&  {-0.3203$\pm$0.0008}   &{-0.3263$\pm$0.0008}&  {-0.2460$\pm$0.0007}   &{-0.2480$\pm$0.0007} \\
{October  }&  {-0.3179$\pm$0.0006}   & {-0.3222$\pm$0.0007} &  {-0.2505$\pm$0.0006}   &   {-0.2525$\pm$0.0006}   \\
{November }&   {-0.3130$\pm$0.0010}   &{-0.3178$\pm$0.0010} &  {-0.2469$\pm$0.0011}   &  {-0.2507$\pm$ 0.0011}  \\   \hline \hline 
 \end{tabular} }

\end{table}

\begin{wrapfigure}{r}{0.55\linewidth}
\centering
 \includegraphics[width = 0.9\textwidth]{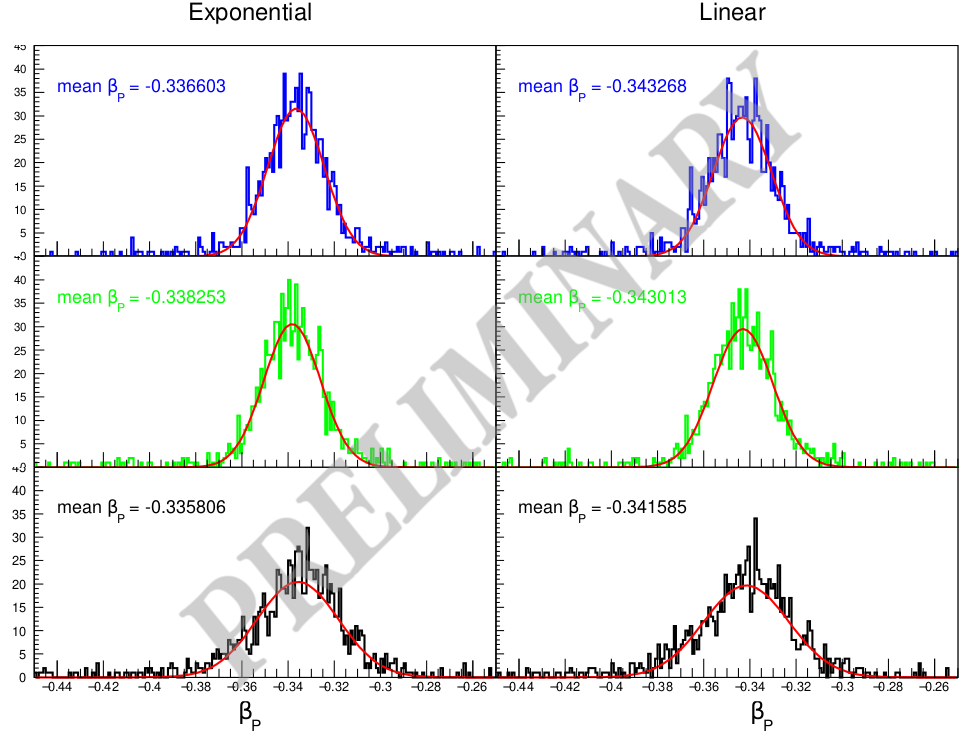}
 \caption{{\scriptsize  \it \label{betall} Distribution of the pressure coefficient $\beta_P$ for all PMTs for the months of \textcolor{blue}{September}, \textcolor{green}{October}, and November. the plots in left side is for exponential method, and that in right sides are by linear approximation method. Mean value of distribution is given along with the figures.}}
\end{wrapfigure}

Similar analyses were carried out for all the single PMT $R_{1}$ rates corresponding to the 1180 PMTs as well as the multiplicity rates $R_{M2}$, $R_{M3}$, \& $R_{M4}$ corresponding to the 295 tanks of HAWC.  These analyses were carried out using data from the months of September, October, and November  of 2016. The pressure coefficients $\beta_P$ were obtained using both the exponential and linear approximations for these months.  The distribution of $\beta_P$ of $R_{1}$ obtained using both methods for these months are shown in figure \ref{betall} and the mean value of $\beta_P$ of $R_{1}$, $R_{M2}$, $R_{M3}$, \& $R_{M4}$ for each month is given in the Table \ref{tbeta}.  The estimated $\beta_P$ for the three months are consistent within the statistical accuracy of the experiment. 


\section{Pressure correction}

\begin{figure}[h]
\begin{floatrow}
\ffigbox{%
\includegraphics[width = 0.75\columnwidth]{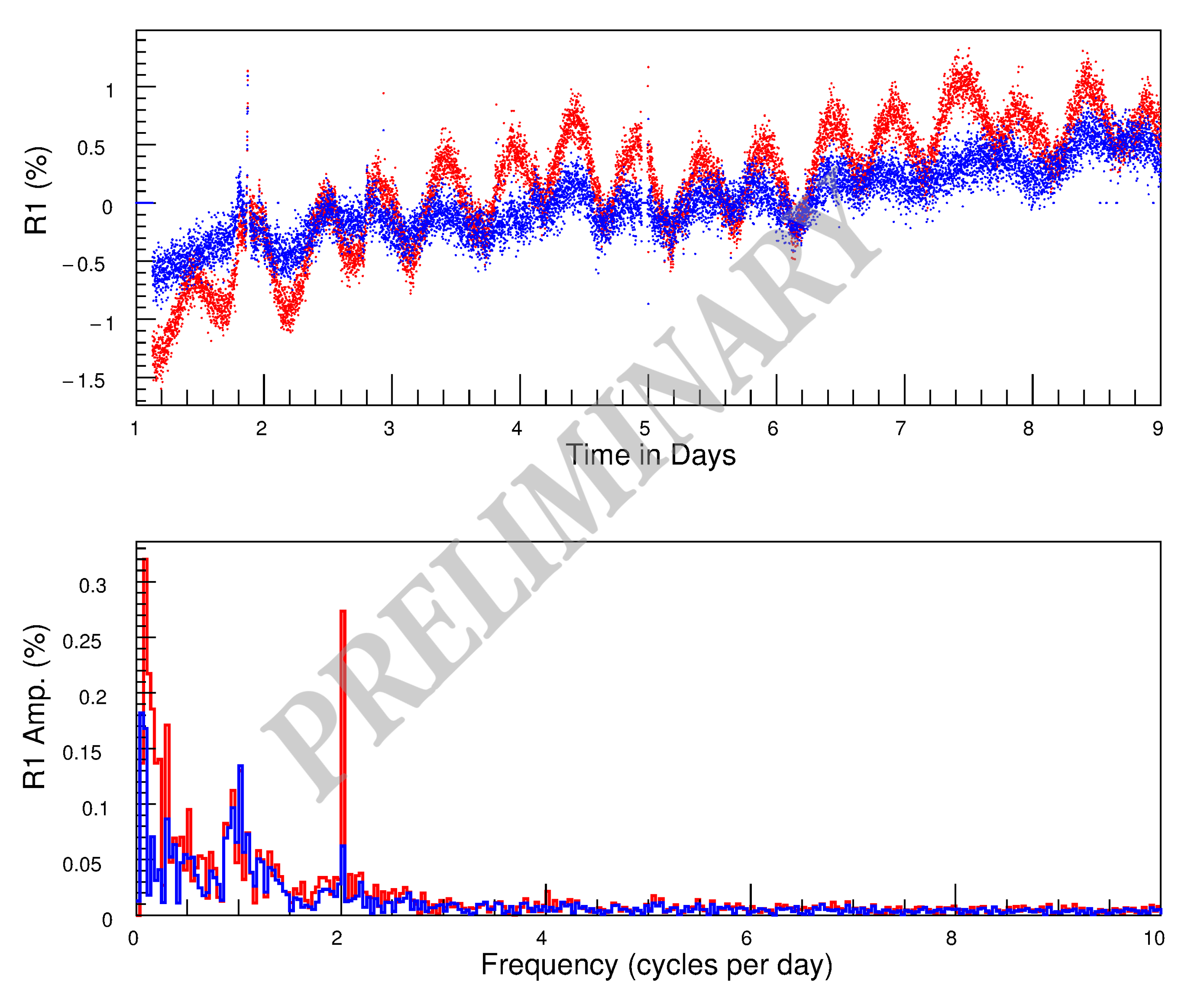}
}{%
 \caption{{\scriptsize  \it \label{corR} Top panel shows the $R_{1}$, the one in red is $R_{1}$ before correction and in blue is the same after pressure correction. The bottom panel shows the power spectra of $R_{1}$, before pressure correction in red and after correction in blue.}}
}
\ffigbox{%
\includegraphics[width = 1.0\columnwidth]{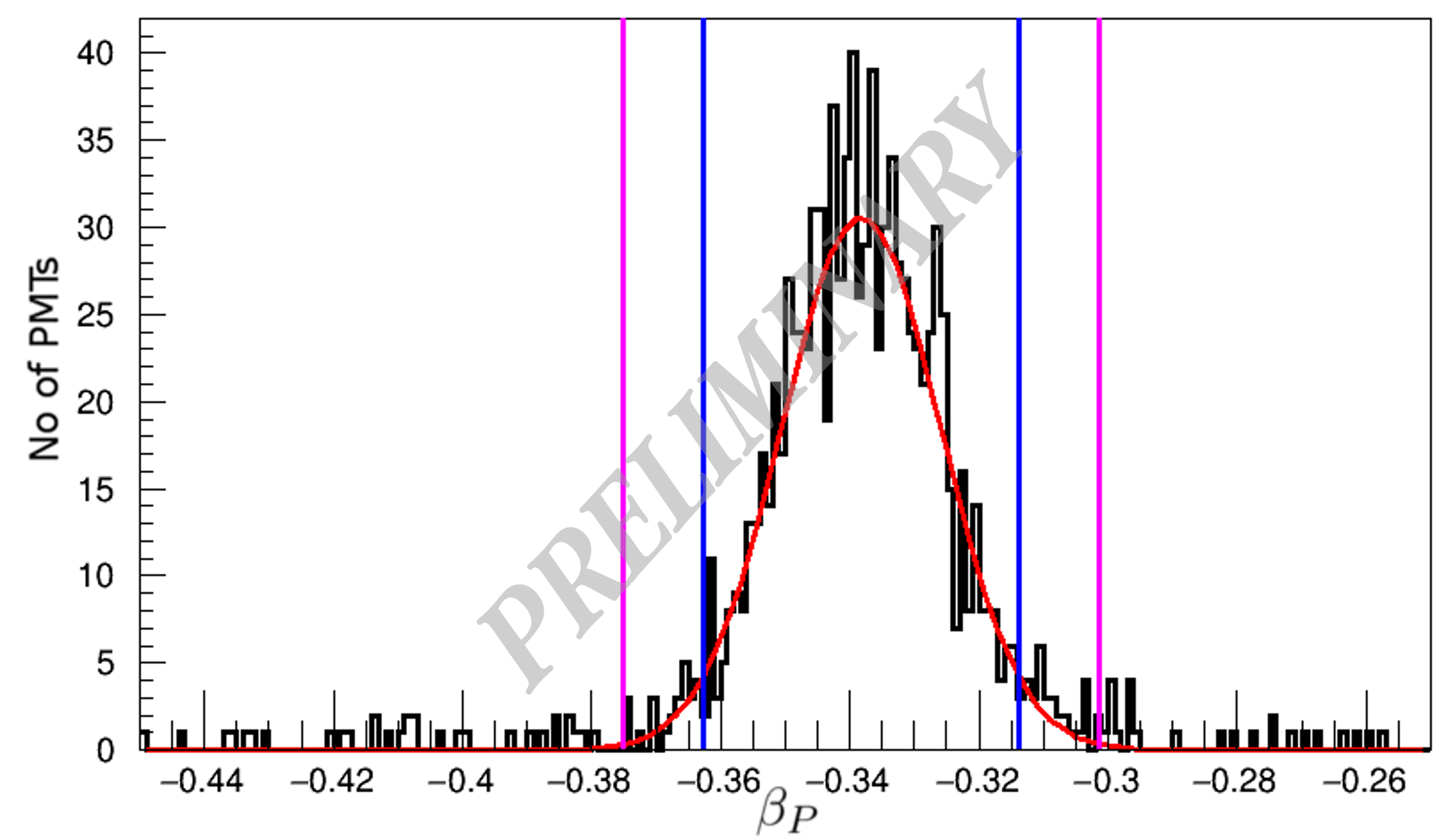}

}{%
 \caption{{\scriptsize  \it \label{Nbp} Distribution of $\beta_P$ for the month of November, the blue vertical lines corresponds to the 2$\sigma$ cut off range and the magenta is the same for 3$\sigma$} }
}
\end{floatrow}
\end{figure}

The mean value of $\beta_P$ was estimated from three months of data using both methods (exponential method $\beta_P$ = -0.337 \%/hPa, linear method $\beta_P$ = -0.343 \%/hPa). The $\chi^2$ for both the methods for each PMT were calculated. The mean $\chi^2$ for linear method was $5.7 \times 10^{-4}$ and that for exponential method was $2.7 \times 10^{-3}$.  It has to be noted that the value of $\beta_P$ for both the methods are consistent within the statistical accuray of the experiment, but considering the best fit for the liear method with the lower $\chi^2 $ value we stick with this method for the presure correction of HAWC scalar system.  The rates $R_{1}$ before and after pressure correction are shown in the first panel of the figure \ref{corR}.  For a comparison, the FFT was applied to the corrected $R_{1}$. The resulting power spectra are shown in the second panel of figure \ref{corR}. It is clear from the figure that the amplitude of 2 cpd was reduced drastically which was mainly due to the pressure component, whereas the amplitudes corresponding to other frequencies were almost unaffected.   Pressure correction were applied to the multiplicities $R_{M2}$, $R_{M3}$ and $R_{M4}$ using the same method.

\section{Checking the health of PMTs and Tanks}
 The cut-off rigidity of vertical protons for HAWC is 7.9 GeV \cite{lar13}, and the median rigidity for the TDC-scaler multiplicities varies within range of $\sim 41 - 45$ GeV. Most of the pions and kaons produced by these low energy protons decay well before they reach Earth because of  short lifetime, hence the modulation observed in TDC-scaler rate by the atmospheric pressure is mainly due to the decay of muons in the atmosphere, and this is well evidenced by the negative correlation observed.   The pressure modulation is a purely physical phenomenon due to a property of the atmosphere and hence will be independent of the detector, thus the value of the pressure coefficient $\beta_P$ for the 1180 PMTs should be similar, any deviation of $\beta_P$ can be due to a malfunction of that particular PMT or its associated components. The distribution of $\beta_P$ for the month of October is shown in the figure \ref{Nbp}. The distribution of $\beta_P$ has a Gaussian nature and the sigma for this Gaussian fit was $\sigma ~ = 1.29 \times 10^{-2}$, the vertical blue and magenta lines show the cut-off range of 2 and 3 $\sigma$. We have classified the health of these PMTs and tanks depending up on their $\beta_P$ value and are shown in the Table \ref{lpmt}. 
The value of $\beta_P$ and its deviation from the mean value gives us a quantitative measure of how well a PMT is functioning in its normal gain mode. With this classification, we chose the PMTs within a 3 $\sigma$ range to include in our analysis of solar modulations. This correction will improve the accuracy of the measurement of GCR modulation by reducing the systematic errors and will make the data more suitable for the solar modulation studies. The smilar selection process were applied to the tanks for the multiplicity rates.


\begin{table}
{\footnotesize
\begin{tabular}{l|ccc|ccc|ccc|ccc}\hline \hline 
&\multicolumn{3}{c}{PMT Rate R1} & \multicolumn{3}{c}{Multiplicity M2} & \multicolumn{3}{c}{Multiplicity M3} & \multicolumn{3}{c}{Multiplicity M4}\\   \hline 
Month & Sep & Oct & Nov & Sep & Oct & Nov  & Sep & Oct & Nov  &  Sep & Oct & Nov  \\ \hline 
Good  & 1027 & 1005 & 999 & 260 & 245 & 226 & 266 & 255 & 263 & 256 & 240 & 250 \\
$3\sigma$ to $ 5\sigma$ & 40 & 24 & 40 & 15 & 11 & 9 & 4 & 7 & 8 & 8 & 9 & 8 \\
Above $ 5 \sigma$ &  103 & 142 &  128 &  18 & 38 &  58 &  23 & 32 &  22  &  24 & 37 &  29\\
No-data & 10 &  9 & 13 & 2 &  1 & 2  & 2 &  1 & 2 & 7 &  9 & 8 \\ \hline \hline 
\end{tabular}
 \caption{{\scriptsize  \it \label{lpmt} Classification of PMTs and Tanks }}}
\end{table}

\section{Summary} 

The study of solar modulations of GCRs is an excellent tool to identify space weather transients. However, an accurate estimation of the pressure coefficient $\beta_P$ and a suitable correction method to remove the pressure induced modulations from the data is required to perform the study of the solar-induced phenomena. The correlation of atmospheric pressure and the observed secondary particle rate of Earth-based detectors are well known, but the interference of the solar modulations such as solar diurnal anisotropy, and Forbush decrease events complicate this relationship. The observed 12-hour periodic nature of pressure at the HAWC site was exploited to extract the strong anticorrelation between the pressure and the TDC-scaler rate. The usage of FFT and a narrow band filter made it possible for the effective isolation of pressure-induced modulation from the solar modulation to obtain this strong anticorrelation and then an accurate estimation of the pressure coefficient $\beta_P$. The consistent result for September, October and November proves the consitancy and accuracy of this method. The obtained value of $\beta_P$ and its deviation from the estimated mean value were used as a quantitative measure to deduce normal performance of a PMT and tank, which is used to remove the abnormally behaving PMTs and tanks from our further analysis for the study of solar modulations. 

{\it {\bf Acknowledgment :} Alejandro Lara thanks  PASPA-UNAM for his partial support.}

\end{document}